\documentstyle[12pt]{article}
\def\be{\begin{equation}}
\def\ee{\end{equation}}
\def\l{\label}

\def\S{{\cal S}}

\def\W{{\cal W}}

\makeatletter \ifcase\@ptsize \font\teneufm=eufm10
\font\seveneufm=eufm7 \font\fiveeufm=eufm5
\font\teneusm=eusm10 \font\seveneusm=eusm7
\font\fiveeusm=eusm5 \or \font\teneufm=eufm10 scaled
\magstephalf \font\seveneufm=eufm7 \font\fiveeufm=eufm5
\font\teneusm=eusm10 scaled \magstephalf
\font\seveneusm=eusm7 \font\fiveeusm=eusm5 \or
\font\teneufm=eufm10 scaled \magstep1 \font\seveneufm=eufm7
\font\fiveeufm=eufm5 \font\teneusm=eusm10 scaled \magstep1
\font\seveneusm=eusm7 \font\fiveeusm=eusm5 \fi

\newfam\eufmfam \newfam\eusmfam \textfont\eufmfam=\teneufm
\scriptfont\eufmfam=\seveneufm
\scriptscriptfont\eufmfam=\fiveeufm
\textfont\eusmfam=\teneusm \scriptfont\eusmfam=\seveneusm
\scriptscriptfont\eusmfam=\fiveeusm

\def\frak{\ifmmode\let\next\frak@\else
 \def\next{\errmessage{Use \string\frak\space only in math
 mode}}\fi\next} \def\frak@#1{{\frak@@{#1}}}
 \def\frak@@#1{\fam\eufmfam#1} 
 \def\sh{\ifmmode\let\next\sh@\else
 \def\next{\errmessage{Use \string\sh\space only in math
 mode}}\fi\next} \def\sh@#1{{\sh@@{#1}}}
 \def\sh@@#1{\fam\eusmfam#1}

\ifcase\@ptsize \font\tenmsa=msam10 \font\sevenmsa=msam7
 \font\fivemsa=msam5 \font\tenmsb=msbm10
 \font\sevenmsb=msbm7 \font\fivemsb=msbm5 \or
 \font\tenmsa=msam10 scaled \magstephalf
 \font\sevenmsa=msam7 \font\fivemsa=msam5
 \font\tenmsb=msbm10 scaled \magstephalf
 \font\sevenmsb=msbm7 \font\fivemsb=msbm5 \or
 \font\tenmsa=msam10 scaled \magstep1 \font\sevenmsa=msam7
 \font\fivemsa=msam5 \font\tenmsb=msbm10 scaled \magstep1
 \font\sevenmsb=msbm7 \font\fivemsb=msbm5 \fi

\newfam\msafam \newfam\msbfam \textfont\msafam=\tenmsa
\scriptfont\msafam=\sevenmsa
\scriptscriptfont\msafam=\fivemsa \textfont\msbfam=\tenmsb
\scriptfont\msbfam=\sevenmsb
\scriptscriptfont\msbfam=\fivemsb

\def\Bbb{\ifmmode\let\next\Bbb@\else
 \def\next{\errmessage{Use \string\Bbb\space only in math
 mode}}\fi\next} \def\Bbb@#1{{\Bbb@@{#1}}}
 \def\Bbb@@#1{\fam\msbfam#1} \def\hexnumber@#1{\ifnum#1<10
 \number#1\else \ifnum#1=10 A\else\ifnum#1=11
 B\else\ifnum#1=12 C\else \ifnum#1=13 D\else\ifnum#1=14
 E\else\ifnum#1=15 F\fi\fi\fi\fi\fi\fi\fi}
 \def\msa@{\hexnumber@\msafam} \def\msb@{\hexnumber@\msbfam}
 \mathchardef\square="0\msa@03

\makeatother
 \newcommand{\RR}{{\Bbb R}}

\begin{document}
\begin{titlepage}

\rightline{UMN-TH-1721-98-TPI-MINN-98/18, DFPD97/TH/58}
\rightline{September 1998, \tt hep-th/9809126}

\vspace{0.133cm}

\begin{center}

{\Large \bf Equivalence Principle: Tunnelling, Quantized Spectra}

\vspace{.333cm}

{\Large \bf and Trajectories from the Quantum HJ Equation}

\vspace{.6333cm}

{\large Alon E. Faraggi$^{1}$ $\,$and$\,$ Marco Matone$^{2}$\\}
\vspace{.2in}
{\it $^{1}$ Department of Physics \\
University of Minnesota, Minneapolis MN 55455, USA\\
        e-mail: faraggi@mnhepo.hep.umn.edu\\}
\vspace{.02in}
{\it $^{2}$ Department of Physics ``G. Galilei'' -- Istituto
                Nazionale di Fisica Nucleare\\
        University of Padova, Via Marzolo, 8 -- 35131 Padova, Italy\\
   e-mail: matone@padova.infn.it\\}

\end{center}

\vspace{.233cm}

\centerline{\large \bf Abstract}

\vspace{.233cm}

A basic aspect of the recently proposed approach to quantum mechanics
is that no use of any axiomatic interpretation of the wave function is made.
In particular, the quantum potential turns out to be an intrinsic potential
energy of the particle, which, similarly to the relativistic rest energy,
is never vanishing. This is related to the tunnel effect, a consequence of
the fact that the conjugate momentum field is real even in the classically
forbidden regions. The quantum stationary Hamilton--Jacobi equation is defined
only if the ratio $\psi^D/\psi$ of two real linearly independent solutions of
the Schr\"odinger equation, and therefore of the trivializing map, is a local
homeomorphism of the extended real line into itself, a consequence of the
M\"obius symmetry of the Schwarzian derivative. In this respect we prove a
basic theorem relating the request of continuity at spatial infinity of
$\psi^D/\psi$, a consequence of the $q\longleftrightarrow q^{-1}$ duality
of the Schwarzian derivative, to the existence of $L^2(\RR)$ solutions of
the corresponding Schr\"odinger equation. As a result, while in the
conventional approach one needs the Schr\"odinger equation with the $L^2(\RR)$
condition, consequence of the axiomatic interpretation of the wave function,
the equivalence principle by itself implies a dynamical equation that does not
need any assumption and reproduces both the tunnel effect and energy
quantization.

\vspace{.633cm}



\end{titlepage}
\setcounter{footnote}{0}
\renewcommand{\thefootnote}{\arabic{footnote}}

\newpage

Let us consider a one--dimensional stationary system of energy $E$ and
potential $V$ and set $\W\equiv V(q)-E$. Let us denote by $\S_0$ the quantum
analogue of the Hamilton characteristic function, also called reduced action.
This function satisfies the Quantum Stationary Hamilton--Jacobi Equation (QSHJE)
\be
{1\over 2m}\left({\partial\S_0(q)\over \partial q}\right)^2+V(q)-E
+{\hbar^2\over 4m}\{\S_0,q\}=0,
\l{1Q}\ee
that in \cite{1} was uniquely derived from the equivalence principle.
It states that

\vspace{.333cm}

\noindent
{\it For each pair $\W^a,\W^b$, there is a transformation
$q^a\longrightarrow q^b=v(q^a)$, such that}
\be
\W^a(q^a)\longrightarrow {\W^a}^v (q^b)=\W^b(q^b).
\l{equivalence}\ee

\vspace{.333cm}

This principle implies the existence of the trivializing map \cite{1}
\be
q\longrightarrow q^{0} =\S_0^{{0}^{\;-1}}\circ \S_0(q),
\l{9thebasicidea}\ee
reducing the system with a given $\W$ to the one corresponding to
$\W^0\equiv 0$.

If $(\psi^D,\psi)$ is a pair of real linearly independent solutions
of the Schr\"odinger equation, then we have
\be
e^{{2i\over \hbar}\S_0}=e^{i\alpha}{w+i\bar\ell\over w-i\ell},
\l{KdT3}\ee
where $w=\psi^D/\psi$, and ${\rm Re}\,\ell\ne 0$.

A basic property of the conjugate momentum $p=\partial_q\S_0$, is that it is
real even in the classically forbidden regions \cite{1}\cite{Floyd}. This fact
is an important check in considering the trajectories described by (\ref{1Q}).
To understand this, observe that in the conventional formulation of quantum
mechanics the tunnel effect is a consequence of the axiomatic interpretation of
the wave function $\psi$. Actually, the fact that it describes the probability
amplitude of finding the particle in the interval $[q,q+dq]$, implies that the
tunnel effect simply arises in the cases in which $\psi$ is not identically
zero in the classically forbidden regions. In the case at hand, there is no need
for any axiomatic interpretation, it is just a consequence of reality of the
conjugate momentum in the classically forbidden regions. This result suggests
considering if the theory reproduces the other fundamental aspect of quantum
mechanics, namely quantization of the energy spectra. This would be a basic
check as also energy quantization is strictly related to the axiomatic
intepretation of the wave function. We will see that this is actually the case.

The properties of the Schwarzian derivative imply that the QSHJE is well
defined if and only if \cite{5}
\be
w\ne cnst, \; w\in C^2(\hat\RR),\; and\; \partial_q^2w\;
differentiable \; on \; \hat\RR ,
\l{ccnn}\ee
and $w(q)$ is locally invertible $\forall q\in \hat {\RR}$, where $\hat {\RR}
\equiv \RR\cup\{\infty\}$ denotes the extended real line. In particular, this
implies the following joining condition at spatial infinity
\be
w(-\infty)=\left\{ \begin{array}{ll} w(+\infty), & for \, w(-\infty)\ne \pm
\infty,\\ -w(+\infty), & for\, w(-\infty)=\pm\infty. \end{array}\right.
\l{specificandoccnn}\ee
It follows that $w$, and therefore the trivializing map, is a local
homeomorphism of $\hat {\RR}$ into itself.

Let us start proving a result concerning the energy spectra. If $V(q)>E$,
$\forall q\in {\RR}$, then, as we will see, there are no solutions such that
the ratio of two linearly independent solutions of the Schr\"odinger equation
corresponds to a local homeomorphism of $\hat {\RR}$ into itself. The fact that
this is an unphysical situation can be also seen from the fact that the case
$V>E$, $\forall q\in {\RR}$, has not classical limit. Therefore, if $V(q)>E$
both at $-\infty$ and $+\infty$, a physical situation requires that there are
at least two points in which $V-E=0$. More generally, if the potential is not
continuous, we should have at least two turning points for which $V(q)-E$
changes sign. Let us denote by $q_-$ ($q_+$) the lowest (highest) value of the
turning points. In the following we will prove the following basic fact

\vspace{.333cm}

{\it If}
\be
V(q)-E\geq \left\{\begin{array}{ll} P_-^2 >0, & q<q_-,\\ 
P_+^2 >0, & q> q_+,\end{array}\right.
\l{asintoticopiumeno}\ee
{\it then the ratio $w=\psi^D/\psi$ is a local homeomorphism of $\hat{\RR}$
into itself if and only if the corresponding Schr\"odinger equation admits an
$L^2({\RR})$ solution.}

\vspace{.333cm}

Note that by (\ref{asintoticopiumeno}) we have
\be
\int^{-\infty}_{q_-}dx \kappa(x)=-\infty, \quad
\int^{+\infty}_{q_+}dx \kappa(x)=+\infty,
\l{divergono}\ee
where
\be
\kappa={\sqrt{2m(V-E)}\over \hbar}.
\l{kappaaaa}\ee

Let us first show that the request that the corresponding Schr\"odinger equation
admits an $L^2({\RR})$ solution is a sufficient condition for the ratio
$w=\psi^D/\psi$ to be continuous at $\pm\infty$. Let us denote by $\psi$ the
$L^2({\RR})$ solution and by $\psi^D$ a linearly independent solution. As we
will see, the fact that $\psi^D$ and $\psi$ are linearly independent imply that
if $\psi\in L^2({\RR})$, then $\psi^D\notin L^2({\RR})$, in particular $\psi^D$
is divergent both at $q=-\infty$ and $q=+\infty$. Let us consider the real ratio
\be
w={A\psi^D+B\psi\over C\psi^D+D\psi},
\l{arbitraryratio}\ee
where $AD-BC\ne 0$. Since $\psi\in L^2({\RR})$,  we have
\be
\lim_{q\longrightarrow\pm\infty}w=\lim_{q\longrightarrow\pm\infty}{A\psi^D+B
\psi\over C\psi^D+D\psi}=\lim_{q\longrightarrow\pm\infty}
{A\psi^D\over C\psi^D}={A\over C},
\l{arbitygvxy}\ee
that is $w(-\infty)=w(+\infty)$. In the case in which $C=0$ we have
\be
\lim_{q\longrightarrow \pm \infty}w=\lim_{q\longrightarrow \pm \infty}
{A\psi^D\over D\psi}=\pm \epsilon\cdot\infty,
\l{arbitygvxyconczero}\ee
where $\epsilon=\pm 1$. The fact that $\lim_{q\to \pm \infty}{A\psi^D\over
D\psi}$ is divergent follows from the mentioned properties of $\psi^D$ and
$\psi$. It remains to understand the fact that if the limit is $-\infty$ at
$q=-\infty$, then this limit is $+\infty$ at $q=+\infty$, and vice versa.
This can be understood by observing that
\be
{\psi^D(q)\over \psi(q)}=c\int^q_{q_0}dx\psi^{-2}(x)+d,
\l{oqqw}\ee
for some real constants $c\ne 0$ and $d$. Now, since $\psi\in L^2({\RR})$, it
follows that $\psi^{-1}\not\in L^2({\RR})$. In particular, $\int^{+\infty}_{q_0
}dx\psi^{-2}(x)=+\infty$ and $\int^{-\infty}_{q_0}dx\psi^{-2}(x)=-\infty$, so
that $\psi^D(-\infty)/\psi(-\infty)=-\epsilon\cdot\infty=-\psi^D(\infty)/\psi
(\infty)$ where $\epsilon={\rm sgn}\, c$.

We now show that the existence in the case (\ref{asintoticopiumeno}) of an
$L^2({\RR})$ solution of the Schr\"odinger equation is a necessary condition
for the ratio $w=\psi^D/\psi$ to be continuous at $\pm\infty$. To this end we
recall that if $V(q)-E\geq P_+^2 >0$, $q>q_+$, then as $q\longrightarrow
+\infty$, we have ($P_+>0$)

\begin{itemize}
\item[{\bf --}]{There is one solution of the Schr\"odinger equation, defined up
to a multiplicative constant, that vanishes at least as $e^{-P_+q}$.}
\item[{\bf --}]{Any other solution diverges at least as $e^{P_+q}$.} 
\end{itemize}

The proof of this fact is based on Wronskian arguments and can be found in
Messiah's book \cite{Messiah}. The above result extends also to the case in
which $V(q)-E\geq P_-^2 >0$, $q<q_-$. In particular, as $q\longrightarrow
-\infty$, we have ($P_->0$)

\begin{itemize}
\item[{\bf --}]{There is one solution of the Schr\"odinger equation,
defined up to a multiplicative constant, that vanishes at least as $e^{P_-q}$.}
\item[{\bf --}]{Any other solution diverges at least as $e^{-P_-q}$.} 
\end{itemize}

These properties imply that if there is a solution of the Schr\"odinger
equation in $L^2({\RR})$, then any solution is either in $L^2({\RR})$ or
diverges both at $-\infty$ and $+\infty$. Let us show that the possibility that
a solution vanishes only at one of the two spatial infinities is excluded.
Suppose that, besides the $L^2({\RR})$ solution, which we denote by $\psi_1$,
there is a solution $\psi_2$ which is divergent only at $+\infty$. On the other
hand, the above properties show that there exists also a solution which is
divergent at $-\infty$. Let us denote by $\psi_3$ this solution. Since the
number of linearly independent solutions of the Schr\"odinger equation is two,
we have
\be
\psi_3=A\psi_1+B\psi_2,
\l{cconnt}\ee
for some constants $A$ and $B$. However, since $\psi_1$ vanishes both at
$-\infty$ and $+\infty$, we have that (\ref{cconnt}) cannot be satisfied unless
$\psi_2$ and $\psi_3$ are divergent both at $-\infty$ and $+\infty$.
This fact and the above properties imply the following

\vspace{.333cm}

{\it If the Schr\"odinger equation has an $L^2({\RR})$ solution,
then any solution has one of the following two possible asymptotic behaviors}

\begin{itemize}
\item[{\bf --}]{Vanishes both at $-\infty$ and $+\infty$
at least as $e^{P_-q}$ and $e^{-P_+q}$ respectively.}
\item[{\bf --}]{Diverges both at $-\infty$ and $+\infty$
at least as $e^{-P_-q}$ and $e^{P_+q}$ respectively.}
\end{itemize}

\vspace{.333cm}

Similarly, we have

\vspace{.333cm}

{\it If the Schr\"odinger equation does not admit an $L^2({\RR})$ solution, 
then any solution has one of the following three possible asymptotic behaviors}

\begin{itemize}
\item[{\bf --}]{Diverges both at $-\infty$ and $+\infty$
at least as $e^{-P_-q}$ and $e^{P_+q}$ respectively.}
\item[{\bf --}]{Diverges at $-\infty$ at least as $e^{-P_-q}$
and vanishes at $+\infty$
at least as $e^{-P_+q}$.}
\item[{\bf --}]{Vanishes at $-\infty$ at least as $e^{P_-q}$
and diverges at $+\infty$
at least as $e^{P_+q}$.}
\end{itemize}

\vspace{.333cm}

\noindent
Let us consider the ratio $w=\psi^D/\psi$ in the latter case.
Since any different choice of linearly independent
solutions of the Schr\"odinger equation corresponds to a M\"obius
transformation of $w$, we can choose\footnote{Here by $\sim$ we mean
that $\psi^D$ and $\psi$ either diverge or vanish ``at least as".}
\be
\psi^D_{\;\stackrel{\sim}{q\longrightarrow -\infty}\;} a_-e^{P_-q},\qquad\qquad
\psi^D_{\;\stackrel{\sim}{q\longrightarrow +\infty}\;} a_+e^{P_+q},
\l{psiddi340}\ee
and
\be
\psi_{\;\stackrel{\sim}{q\longrightarrow -\infty}\;} b_-e^{-P_-q},\qquad\qquad
\psi_{\;\stackrel{\sim}{q\longrightarrow +\infty}\;} b_+e^{-P_+q}.
\l{psii340}\ee
Their ratio has the asymptotics
\be
{\psi^D\over\psi}{}_{\;\stackrel{\sim}{q\longrightarrow -\infty}\;}
c_-e^{2P_-q}\longrightarrow 0
,\qquad\qquad
{\psi^D\over\psi}{}_{\;\stackrel{\sim}{q\longrightarrow +\infty}\;}
c_+e^{2P_+q} \longrightarrow \pm\infty,
\l{psiddisupsi340}\ee
so that $w$ cannot satisfy the continuity condition
(\ref{specificandoccnn}) at $\pm\infty$.

The above results imply that the quantized spectrum one obtains from the
conventional approach to quantum mechanics arises as a consequence of the
equivalence principle. Actually, even in the conventional approach the
quantized spectrum and its structure arose by the condition that the values of
$E$ satisfying (\ref{asintoticopiumeno}) should correspond to a Schr\"odinger
equation having an $L^2(\RR)$ solution. Then, all the standard results on the
quantized spectrum are reproduced in our formulation.

Let $n$ be the index $I[q^0]$ of the trivializing map. This is the number of
times $q^0$ spans $\hat{\RR}$ while $q$ spans $\hat{\RR}$. In other words, $n$
is the index of the covering associated to the trivializing map.
Since $q^0$ and $w$ are related by a M\"obius transformation \cite{5},
we have that the index of $q$ and $w$ coincide
\be
I[q^0]=I[w].
\l{qzeroew}\ee
Another property of the trivializing map is that its index depends on $\W$ but
not on the specific M\"obius state. The M\"obius states are the states with the
same $\W$ but with different values of the constant $\ell$ (see \cite{5} for
related aspects). Observe that since $p=\partial_q\S_0$ does not vanish for
finite values of $w$, it follows that $I[q^0]$ coincides with the number of
zeroes of $w$. This aspect may be also understood by recalling a Sturm theorem
about second--order differential equations. The theorem states that given two
linearly independent solutions $\psi^D$ and $\psi$ of the equation $\psi''(q)=
K(q)\psi(q)$, between any two zeroes of $\psi^D$ there is one zero of
$\psi$.\footnote{Observe that this can be seen as a sort of duality between
$\psi^D$ and $\psi$. In this context we note that, while in the conventional
approach one usually selects the wave function which is a particular solution
of the Schr\"odinger equation, $\S_0$ and $p$ contain both $\psi^D$ and $\psi$.}
This theorem, and the condition that the values of $E$ satisfying 
(\ref{asintoticopiumeno}) should correspond to a Schr\"odinger equation
having an $L^2(\RR)$ solution guarantees local homeomorphicity of the
trivializing map. It remains to understand the case in which $V(q)>E$,
$\forall q\in {\RR}$. We already noticed that, since there is not the classical
limit in this case, these solutions are not admissible ones. We now show that
these solutions do not satisfy the continuity condition for $w$ at $\pm\infty$.
To see this it is sufficient to note that if $\psi$ decreases as
$q\longrightarrow -\infty$, then by $\psi''/\psi=4m\W/\hbar^2>0$, $\forall
q\in {\RR}$, it follows that $\psi$ is always convex, $\psi\not\in L^2(\RR)$.
The absence of turning points does not modify the essence of the above
conclusions and if $V(q)>E$, $\forall q\in {\RR}$, then the ratio $\psi^D/\psi$
is discontinuous at $\pm\infty$. As an example, let us consider the equation
\be
{\hbar^2\over 2m}\partial_q^2\psi = a^2\psi.
\l{esempioill2}\ee
Any pair of linearly independent solutions has the form
\be
\psi^D=Ae^{aq}+Be^{-aq},\qquad \psi=Ce^{aq}+De^{-aq},
\l{labella}\ee
where
\be
AD-BC\ne 0.
\l{labelloTIDEilmitico}\ee
Their ratio
\be
{\psi^D\over \psi} ={Ae^{2aq}+B\over Ce^{2aq}+D},
\l{rapportino}\ee
has the asymptotics
\be
\lim_{q\longrightarrow -\infty}{\psi^D\over \psi}={B\over D},\qquad
\lim_{q\longrightarrow +\infty}{\psi^D\over \psi}={A\over C},
\l{VxFTg}\ee
so that by (\ref{labelloTIDEilmitico}) neither the case $w(-\infty)=finite=
w(+\infty)$, nor $w(-\infty)=-w(+\infty)=\pm\infty$ can occur.

We now consider the cases of the potential well and of the simple and double 
harmonic oscillators. We will explicitly see that in the case one considers the
energy values for which the corresponding Schr\"odinger equation has not $L^2(
\RR)$ solutions, the ratio $\psi^D/\psi$ has a discontinuity at
spatial infinity.

Let us consider the potential well
\be
V(q)=\left\{\begin{array}{ll} 0, & |q|\leq L, \\ V_0, & |q|> L,
\end{array}\right.
\l{Vu1}\ee
and set
\be
k={\sqrt{2mE}\over \hbar}, \qquad \kappa={\sqrt{2m(V_0-E)}\over \hbar}.
\l{I91d}\ee
According to (\ref{KdT3}), in order to determine $\S_0$, and therefore 
to solve the dynamical problem, we have to find two real linearly independent
solutions of the Schr\"odinger equation.
Since the potential is even, we can choose solutions of definite parity. We have
\be
\psi=k^{-1}\cdot \left\{ \begin{array}{ll} -\alpha \exp [\kappa(q+L)] -\beta
\exp [-\kappa(q+L)], & q<-L,\\ \sin(kq), & |q|\leq L,\\ \alpha \exp [-\kappa(
q-L)] +\beta \exp [\kappa(q-L)] , & q>L, \end{array}\right.
\l{casogenerale1}\ee
where for any $E\geq 0$
\be
\alpha={1\over 2}\sin (kL)-{k\over 2\kappa}\cos (kL) ,\qquad
\beta={1\over 2}\sin (kL)+{k\over 2\kappa}\cos (kL).
\l{casogeneraledue}\ee
For the dual solution, we have
\be
\psi^D=\left\{ \begin{array}{ll} \gamma \exp [\kappa(q+L)] +\delta \exp
[-\kappa(q+L)], & q<-L,\\ \cos(kq), & |q|\leq L,\\ \gamma \exp [-\kappa(q-L)]
+\delta \exp [\kappa(q-L)] , & q>L, \end{array}\right.
\l{casogenerale3}\ee
where
\be
\gamma={1\over 2}\cos (kL)+{k\over 2\kappa}\sin (kL) ,\qquad
\delta= {1\over 2}\cos (kL)-{k\over 2\kappa}\sin (kL).
\l{casogeneralequattro}\ee

The ratio of the solutions is given by
$$
{\psi^D\over \psi} =
$$
\be
k\cdot \left\{ \begin{array}{ll} -(\gamma \exp [\kappa(q+L)] +\delta \exp
[-\kappa(q+L)])/ (\alpha \exp [\kappa(q+L)]+\beta \exp [-\kappa(q+L)]), & q<-L,
\\ \cot (kq), & |q|\leq L,\\ (\gamma \exp [-\kappa(q-L)] +\delta \exp [\kappa(
q-L)])/ (\alpha \exp [-\kappa(q-L)] +\beta \exp [\kappa(q-L)]), & q>L,
\end{array}\right.
\l{casogenerale5}\ee
whose asymptotic behavior is
\be
\lim_{q\longrightarrow \pm\infty}{\psi^D\over \psi} =\pm {\delta\over \beta}k.
\l{limitidoversisplendido2generale}\ee

Continuity at $\pm\infty$ implies that either
\be
\beta=0,
\l{ariprovace}\ee
so that $w(-\infty)=-{\rm sgn}\, \delta \cdot \infty=-w(+\infty)$, or
\be
\delta =0,
\l{toprocnc}\ee
so that $w(-\infty)=0=w(+\infty)$. It is easy to see that 
(\ref{casogeneraledue})(\ref{casogeneralequattro})(\ref{ariprovace}) and
(\ref{toprocnc}) identify the usual quantized spectrum.

Let us now consider the double and simple harmonic oscillators. The Hamiltonian
describing the relative motion of the reduced mass of the double harmonic
oscillator is
\be
H={p^2\over 2m}+{1\over 2}m \omega^2(|q|-q_0)^2.
\l{doppiooscarm}\ee
The reduced action for this system is given by (\ref{KdT3}) with $\psi^D$ and
$\psi$ real linearly independent solutions of the Schr\"odinger equation
\be
\left(-{\hbar^2\over 2m}{\partial^2\over \partial q^2}+
{1\over 2}m\omega^2(|q|-q_0)^2\right)\psi=E\psi.
\l{schrdoparosc}\ee
Let us set
\be
E=\left(\mu+{1\over 2}\right)\hbar\omega,
\l{muiota}\ee
and
\be
z'=\sqrt{2m\omega\over\hbar}(q+q_0),\quad q\leq 0,
\qquad z=\sqrt{2m\omega\over\hbar}(q-q_0), \quad q\geq 0.
\l{zzprimo}\ee
Observe that $z'(-q)=-z(q)$ and that for $q_0=0$, $z'=z=\sqrt{2m\omega\over\hbar}q$,
so that the system reduces to the simple harmonic oscillator.

We stress that since we are considering the equation (\ref{schrdoparosc}) for
arbitrary real $E$, we have that at this stage $\mu$ is an arbitrary real
number.

The Schr\"odinger equation (\ref{schrdoparosc}) is equivalent to
\be
{\partial^2\psi\over \partial {z'}^2}+
\left(\mu+{1\over 2}-{{z'}^2\over 4}\right)\psi=0, \qquad q\leq 0,
\l{schrdoparosc1}\ee
and
\be
{\partial^2\psi\over \partial z^2}+
\left(\mu+{1\over 2}-{z^2\over 4}\right)\psi=0, \qquad q\geq 0.
\l{schrdoparosc2}\ee

For any $\mu$ we have that a solution of (\ref{schrdoparosc2}) is given
by the parabolic cylinder function (see for example \cite{MagnusOberhettinger})
$$
D_\mu(z)=2^{\mu/2}e^{-z^2/4}\left[ {\Gamma(1/2)\over \Gamma[(1-\mu)/2]}
{}_1F_1(-\mu/2;1/2;z^2/2)\right.
$$
\be
\left. +{z\over \sqrt{2}} {\Gamma(-1/2)\over \Gamma(-\mu/2)}
{}_1F_1[(1-\mu)/2;3/2;z^2/2] \right],
\l{cilindroparabolico}\ee
where ${}_1F_1$ is the confluent hypergeometric function
\be
{}_1F_1(a;c;z)=1+{a\over c}{z\over 1!}+{a(a+1)\over c(c+1)}{z^2\over 2!}+
\ldots .
\l{ipergconfl}\ee

We are interested in considering the continuity of the ratio $w=\psi^D/\psi$
at $\pm\infty$. That is, if $w(-\infty)\ne\pm\infty$ we have to impose
$w(-\infty)=w(+\infty)$, while if $w(-\infty)=\pm\infty$, then we should have
$w(-\infty)=-w(+\infty)$. To study the continuity at $\pm\infty$ we need the
behavior of $D_\mu$ for $|z|\gg 1$, $|z|\gg |\mu|$. In the case $\pi/4<\arg z
<5\pi/4$ we have
$$
D_\mu(z){}_{\;\stackrel{\sim}{|z|\gg 1}\;} -{\sqrt{2\pi}\over \Gamma(-\mu)}
e^{\mu\pi i}e^{z^2/4}z^{-\mu-1}\left[1+{(\mu+1)(\mu+2)\over 2  z^2}\right.
$$
$$
\left. +{(\mu+1)(\mu+2)(\mu+3)(\mu+4)\over 2\cdot 4 z^4}+\ldots\right]
$$
\be
+e^{-z^2/4}z^\mu\left[1-{\mu(\mu-1)\over 2 z^2}
+{\mu(\mu-1)(\mu-2)(\mu-3)\over 2\cdot 4  z^4}-\ldots\right],
\l{menoinf}\ee
while for $|\arg z|<3\pi/4$
\be
D_\mu(z){}_{\;\stackrel{\sim}{|z|\gg 1}\;} e^{-z^2/4}z^\mu\left[1-{\mu(\mu-1)
\over 2 z^2}+{\mu(\mu-1)(\mu-2)(\mu-3)\over 2\cdot 4 z^4}-\ldots\right].
\l{piuinf}\ee

A property of the cylinder parabolic function is that even $D_\mu(-z)$ is a
solution of (\ref{schrdoparosc2}). In particular, if $\mu$ is a non--negative
integer, then $D_\mu(z)$ and $D_\mu(-z)$ coincide.\footnote{Note that in the
case in which $\mu$ is a non--negative integer we have $\Gamma^{-1}(-\mu)=0$,
so that in this case the first term in (\ref{menoinf}) cancels.} Let us then
consider the values of $\mu$ different from a non--negative integer. We have
\be
\psi =\left\{ \begin{array}{ll} D_\mu(-z')+cD_\mu(z'),& q\leq 0,\\
{} & {}\\ D_\mu(z)+cD_\mu(-z), & q\geq 0, \end{array}\right.
\l{zummmparidmudiv}\ee
where
\be
c={D_\mu'(-\alpha q_0)\over D_\mu'(\alpha q_0)}.
\l{FrankZappa56}\ee
For any given $\mu$, a linearly independent solution is given by
\be
\psi^D =\left\{ \begin{array}{ll} -D_\mu(-z')-dD_\mu(z'),& q\leq 0,\\
{} & {}\\ D_\mu(z)+dD_\mu(-z), & q\geq 0, \end{array}\right.
\l{zummmdisparidmudiv}\ee
where
\be
d=-{D_\mu(-\alpha q_0)\over D_\mu(\alpha q_0)}.
\l{FrankZappa562}\ee

The ratio
\be
{\psi^D_+\over \psi_-}=\left\{ \begin{array}{ll}-(D_\mu(-z')+dD_\mu(z'))/(
D_\mu(-z')+cD_\mu(z')),& q\leq 0,\\ {} & {}\\ (D_\mu(z)+dD_\mu(-z))/ (D_\mu(z)+
cD_\mu(-z)), & q\geq 0, \end{array}\right.
\l{zummmdsjhdoisparidmudiv}\ee
has the asymptotics behavior
\be
\lim_{q\longrightarrow \pm\infty}{\psi^D\over\psi}=\pm{d\over c}.
\l{dsucabbastanza}\ee
This shows that continuity at $\pm\infty$ is satisfied in the case in which
either $c=0$ or $d=0$, which fix the standard energy quantized spectra (see
also \cite{5}).

Observe that $\psi$ in (\ref{zummmparidmudiv}) and the dual solution
(\ref{zummmdisparidmudiv}) are not linearly independent in the case in which
$\mu=0,1,2,\ldots$. In this case there is always a solution vanishing both at
$-\infty$ and $+\infty$. In the case $q_0=0$, this situation corresponds to
the harmonic oscillator. Generally, for an arbitrary $q_0$ and
$\mu=0,1,2,\ldots$, the solution
$\psi^D$ in (\ref{zummmdisparidmudiv}) is replaced by
\be
\psi^D =\left\{ \begin{array}{ll} -D_n(-z')-d'D_{-n-1}(-iz'),& q\leq 0,\\
{} & {}\\ D_n(z)+d'D_{-n-1}(iz), & q\geq 0,
\end{array}\right.
\l{zummmdisparidmudiv3456}\ee
where now
\be
d'=-{D_n(-\alpha q_0)\over D_n(-i\alpha q_0)}.
\l{FrankZappa28ottobre56}\ee
We note that above we used the parabolic cylinder functions with real argument.
Then, the fact that for $\mu=0,1,2,\ldots$, $D_\mu(z)$ and $D_\mu(-z)$ are not
linearly independent forced us to use $D_\mu(-iz)$. In this context we observe
that $D_\mu(z)$ and $D_\mu(-iz)$ are always linearly independent so that the
dual solution (\ref{zummmdisparidmudiv3456}) can be extended to arbitrary
values of $\mu$.

\vspace{.333cm}

Work supported in part by DOE Grant No.\ DE--FG--0287ER40328 (AEF)
and by the European Commission TMR programme ERBFMRX--CT96--0045 (MM).

\end{document}